\documentclass[twocolumn,superscriptaddress]{revtex4-1}

\usepackage{tikz}

\usetikzlibrary{decorations.pathmorphing}
\usetikzlibrary{decorations.markings}
\usetikzlibrary{positioning}
\usetikzlibrary{fadings,shadings,shapes}
\usetikzlibrary{patterns}
\definecolor{shinycol}{rgb}{0.75, 0.75, .85}
\definecolor{transcol}{rgb}{.85, .85, .9}
\usetikzlibrary{arrows}
\tikzset{cross/.style={cross out, draw, 
         minimum size=2*(#1-\pgflinewidth), 
         inner sep=0pt, outer sep=0pt}}
         
\usepackage{acronym}

\usepackage{subfigure}
\usepackage{amssymb,mathrsfs}
\usepackage{siunitx}
\usepackage{helvet}
\usepackage{hyperref}
\hypersetup{
    colorlinks=true,
    breaklinks=true,
    citecolor=black,
    linkcolor=black,
    menucolor=black,
    urlcolor=black,
}

\begin{document}

\title{Designing Gram-Scale Resonators for Precision Inertial Sensors}

\author{Jonathan J. Carter}
\affiliation{ Max Planck Institute for Gravitational Physics (Albert Einstein Institute), Callinstr. 38, Hannover, Germany}
\affiliation{ Institute for Gravitational Physics of the Leibniz Universit\"at Hannover, Callinstr. 38, Hannover, Germany}

\author{Pascal Birckigt}
\affiliation{Fraunhofer Institute for Applied Optics and Precision Engineering, Albert-Einstein-Str.~7, Jena, Germany}

\author{Oliver Gerberding}
\affiliation{Institut für Experimentalphysik, Universität Hamburg, Luruper Chaussee~149,
 Hamburg, Germany}

\author{Sina M. Koehlenbeck}
\affiliation{ Max Planck Institute for Gravitational Physics (Albert Einstein Institute), Callinstr. 38, Hannover, Germany}
\affiliation{ Institute for Gravitational Physics of the Leibniz Universit\"at Hannover, Callinstr. 38, Hannover, Germany}

\date{\today }

\begin{abstract}
Recent advances in glass fabrication technology have allowed for the development of high-precision inertial sensors in devices weighing in the order of grams. Gram-scale inertial sensors can be used in many applications with tight space or weight requirements. A key element of these devices' performance is the behaviour of a mechanical resonator. We present a detailed study on the design of resonators for such sensors. First, we consider how the mechanical parameters of a resonator couple with an inertial sensor's performance. Then, we look at how to geometrically design resonators to achieve specific mechanical behaviour without undergoing brittle failure. Both analytic tools and finite element analysis are used to this end. We then derive expressions that can be used to optimise the performance of an inertial sensor for a specific sensitive bandwidth. A simple geometry used throughout the field is studied as an example. However, the results are presented in a general form so they can easily be adapted to any required geometry and use case. Ultimately, the results presented here guide the design of gram-scale inertial sensors and will improve the performance of devices that follow them.

\end{abstract}

\maketitle
\acrodef{Q}[$Q$ factor]{Mechanical Quality factor}

\acrodef{TED}{Thermoelastic Damping}
\acrodef{FEA}{Finite Element Analysis}
\acrodef{TTL}{Tilt to Length Coupling}
\acrodef{ASD}{Amplitude Spectral Density}


\section{Introduction}
Gram-scale, precision inertial sensors are a field seeing rapid innovation. \cite{Guzman2014,Gerberding2015,Carter2020a,Hines2020,Nelson2022,Kumanchik2023,Capistran2023,Hines2023} 
These sensors aim to achieve a state-of-the-art performance in compact, vacuum-compatible housings. They achieve high performance through careful design to minimise noise from the thermal noise of the suspension system and interferometric readouts to measure the mechanical motion of the device precisely. Many of these sensors are designed to work alongside sensitive physics experiments, measuring and helping isolate experiments from seismic disturbances. Implementation of such sensors alongside atomic interferometers has been demonstrated, \cite{Richardson2020} and there is a push to integrate such sensors in the control and isolation systems of gravitational wave detectors. \cite{Hines2023,Matichard2015} Sensors of this style are starting to make their way out of the confines of laboratories and seeing more dynamic applications, such as on spacecraft used in geodesy. \cite{Sanjuan2022} 
\par
Inertial sensors need two parts: a means of encoding inertial motion and a means of reading out this motion. 
gram-scale sensors demonstrated so far have vastly different design geometries, properties and scopes to meet. \cite{Guzman2014,Gerberding2015,Carter2020a,Hines2020,Nelson2022,Kumanchik2023,Capistran2023,Hines2023} However, they all have some features in common. Inertial motion is encoded into the motion of a mechanical resonator. An interferometric displacement sensor reads out the motion of this resonator. The devices' small size makes them susceptible to suspension thermal noise, one of the leading noise terms in many designs. \cite{Carter2020a} A high \ac{Q} mode of oscillation
is needed to suppress thermal noise and demands low mechanical bulk loss materials such as fused silica and silicon. \cite{Nelson2022} The \ac{Q} can be understood mechanically as 
\begin{equation}
    Q=2\pi\frac{\rm{Energy \ Stored}}{\rm{Energy\ Dissipated\ per\ Oscillation\ Cycle}},
    \label{eqn:Qwordydef}
\end{equation}
effectively making it a measure of how well energy is stored in the oscillating system.
The mechanical resonator is usually designed with two or more thin bridges between the suspended test masses and the outer frame, which we call the flexures. The flexures usually have a thickness of the order of \SI{100}{\micro\meter} and lengths of millimetres. These flexures then support test masses over 1 gram. Using multiple flexures, the fundamental mode of oscillation can be linear with respect to the sensing axis. Typically, the features that improve the noise performance of these resonators can also lead to brittle fracture of the thin flexures. Compromises have to be made in the design between improving noise performance and surviving operation and transport.
\par
Most successful resonators in this field have been manufactured by an etching-assisted femtosecond laser ablation method first developed by Bellouard et al.\ \cite{Bellouard2012} and made commercially available by FEMTOprint. The technique uses a two-step process, whereby first, any areas to be removed are ``activated" by a high-energy femtosecond pulse laser. Then, the whole sample is bathed in HF acid, leaving only the desired geometry. The method allows for complex geometries to be produced. \cite{Carter2020a,Kumanchik2023}
The drawback of this method is that control of the surface roughness is not well preserved and coating before structuring is not possible.

As the field aims to make increasingly high-performance sensors, optimal design becomes critical. We present a robust approach to the design of the resonators, which are optimised for noise performance. This paper focuses on the design of the resonating part and best practices that minimise a mechanical resonator's thermal noise without suffering brittle fracture while in operation or transport. To do this, we look at loss terms in mechanical resonators and how they couple to noise performance in section \ref{sec:nosie}. Then, in section \ref{sec:DesignofLow}, a study is done on how the parameters that define a flexure geometry couple to the mechanical behaviour of the resonator. Both \ac{FEA} and analytical modelling are undertaken. These results are combined in section \ref{sec:minther}, where the optimum choice for parameters for specific design cases are found. Here, values for a low-frequency fused silica resonator in a laboratory environment are used as an example. General equations and optimums are defined, which can be used in designing resonators for a wide range of applications. We conclude with remarks about implementing these results into designs for inertial sensors.

\section{Noise Sources In Inertial Sensors}

\label{sec:nosie}
An inertial sensor can be viewed as a suspended test mass, which behaves as a simple harmonic resonator in a box. Inertial motion, $X_{\rm{g}}$, the system's input, is then read out as the measurable distance from the edge of the box to the test mass, $\Delta X$, the system's output, by the transfer function
\begin{equation}
    \frac{\Delta X (\omega)}{X_{\rm{g}}(\omega)}=\frac{-\omega^2}{\omega^2-{\omega_0}^2-\frac{i{\omega_0}^2}{Q}},
    \label{eqn:tfseismo}
\end{equation}
where $\omega$ is the angular Fourier frequency and $\omega_0$ is the natural angular frequency of oscillation of the fundamental mode. 
\par
The noise of an inertial sensor can be characterised into two groups:  noise sources that disturb the position of a test mass and noises causing a measurement error of the test mass position. As gram-scale sensors typically have higher resonance frequencies than other high-precision sensors, the motion of the test mass to the system input will be smaller. This mandates a precision readout method. As we focus on resonator design, the effects relevant to us are those that disturb the test mass position. There is much research on low noise readout schemes that are suitable for integration with a low noise resonator to make a complete inertial sensor \cite{Gerberding2015,Cooper2018,Zhou2021,krause2012,Gerberding2015a,Isleif2019,Smetana2022,Yang2020,Smith2009,Zhang2022}.
\par
As a consequence of \ref{eqn:tfseismo}, When the readout noise is white in terms of displacement, it will be white in inertial equivelent displacment units above the $\omega_0$ of the resonator. Below its resonance, it will increase as $1/\omega^2$. Therefore, to widen the low-readout-noise band, high performance inertial sensors commonly use a resonator with a low resonance frequency. Increasing ground motion at low frequencies and needing high \ac{Q} resonators can create dynamic range issues if the inertial sensor is used as a seismometer. These factors often limit suitable readout techniques or a resonator's acceptable values of $\omega_0$.
\par
\subsection{Suspension Thermal Noise}
Suspension thermal noise is the primary source of noise that disturbs the test mass position. It originates from the thermally-driven excitations of the microscopic degrees of freedom of the test mass coupling to test mass motion through the fluctuation-dissipation theorem. It is often the fundamental limit of a design. Equations defining the limits of this noise source have been well defined in several places \cite{saulson1994,Agatsuma2010}.
Suspension thermal noise typically becomes a problem for low frequency sensors.
\par
How thermal noise scales depends on the damping mechanism. 
When the damping is related to internal flexure behaviour, it usually depends on displacement. This is called structural damping. 
The acceleration noise from structural damping is given by  
\begin{equation}
	\tilde{A}(\omega)=\sqrt{\frac{4k_{\rm{b}}T\omega_0^2}{m\omega Q}}.
	\label{eqn:strucNoise}
\end{equation}
Here, the extra $\omega_0/\omega$ term creates a slope of extra noise at low frequency. The factors that make a low noise inertial sensor are already apparent. We need a high mass, low natural frequency, and high \ac{Q}. Large inertial sensors achieve low damping losses using large proof masses with soft suspensions. Gram-scale inertial sensors must compensate for this mass loss to achieve high precision by using high \acp{Q}, typically at least in the order of $10^4$\cite{Guzman2014,Carter2020a,Hines2023}. 
\par
We must, therefore, consider the loss terms dominant in gram-scale resonators and how to minimise them.
\subsubsection{Thermoelastic Damping} 
\label{sec:ted}
\ac{TED} has been well derived and described \cite{Zener1949,lifshitz2000,Norris2005}, and only the relevant results are stated here.
\par
The Zener approximation can estimate \ac{TED} in thin beams, as is the method used by Lifshitz et al.\ \cite{lifshitz2000}. A derivation of a direct solution for the plate geometry similar to thin flexures is presented by Norris and Photiadis \cite{Norris2005}. The key result from this is that the \ac{Q} from \ac{TED} can be approximated as
 \begin{equation}
 	Q^{-1} _ { \mathrm { TED } } \approx \frac { \alpha ^ { 2 } E T} {  C _ { \mathrm { p } } } \frac { \omega \tau } { 1 + \omega ^ { 2 } \tau ^ { 2 } }
 	\label{eqn:ted}
 \end{equation}
 where $\alpha$ is the coefficient of thermal expansion, $E$ is the Youngs modulus,  $\rho$  the density, $C_p$  the volumetric heat capacity at constant pressure, and $\tau$ is the thermal relaxation time. In the case of a thin beam, this is approximated by 
 \begin{equation}
 	\tau = \frac { {d_{\rm{fwidth}}} ^ { 2 } } { \pi ^ { 2 } \chi }
 	\label{eqn:tedrelax}
 \end{equation}
 with $d_{\rm{fwidth}}$ being the flexure thickness, and $\chi$ the thermal diffusivity. Other geometries have been solved, but solutions become increasingly complex \cite{Norris2005,Chadwick1962}. For this reason,  more complex geometries often use \ac{FEA} to solve the effect of thermoelastic damping \cite{Duwel2006,Guo2013}. 
\par
The \ac{Q} has an explicit dependence on frequency. When substituted into \ref{eqn:strucNoise}, the frequency dependence cancels out at frequencies significantly below the thermal relaxation time, and thermal noise becomes flat with respect to frequency. The second key effect is the occurrence of the \ac{TED} peak. It is, therefore, advisable to tune design parameters such that the \ac{TED} peak is outside of sensitive frequency ranges.
\par
\subsubsection{Material, Bulk, and Surface Loss}
\label{sec:matloss}
Another leading loss term is the intrinsic loss of the material. The bulk of any material will have channels in which the energy can be dissipated; typical values for fused silica are of the order $Q^{-1}_{\mathrm{bulk}}=10^{-7}$ at room temperature \cite{Numata2002}. The material's surface will have a less regular structure, and contaminants may be embedded in the structure, increasing mechanical loss along the surface. Typically, this is treated as having a bulk material with the materials intrinsic loss and a surface layer with a lower intrinsic loss, $Q^{-1}_{\mathrm{Surf}}$ penetrating to some depth $d_{\rm{s}}$. Gretarson et al.\ estimated a value of $Q^{-1}_{\mathrm{Surf}}=10^{-5}$, in the case of well handled fused silica fibres \cite{Gretarsson1999}. The value could vary wildly based on treatment, handling, and manufacturing, but it is useful as an estimate for samples kept in a clean environment. 
We use this interpretation to estimate a material loss term of 
\begin{equation}
	Q^{-1}_{\mathrm{{Mat}}} = Q^{-1}_{\rm{Bulk }}+Q^{-1}_{\rm{Surf }} d_{\rm{s}} \frac{\int_{S} \epsilon(\vec{r}) \mathrm{d} S}{\int_{V} \epsilon(\vec{r}) \mathrm{d} V}
    \label{eqn:bulkloss}
\end{equation}
where $\epsilon$ is the strain energy density per unit volume. If a rough estimate is needed, one can reasonably assume that the strain energy is evenly distributed over the oscillating area or can solve using \ac{FEA}. In order to estimate values of $d_{\rm{s}}$, a surface roughness measurement must be made. Surface roughness will depend heavily on manufacturing techniques. As the production methods of FEMTOprint \cite{Bellouard2012} have currently shown the best results \cite{Guzman2014,Hines2023} we take measurements on samples produced in this manner and use that as the estimates for the discussion in Section \ref{sec:minther}.
\par
Fused silica is a standard choice for a room temperature high Q resonator \cite{Guzman2014}. Fused silica is used due to its low bulk loss and comparatively high shear stress to other low-loss glasses \cite{Numata2002}. Numata et al.\ studied the specific bulk losses for several types of fused silica \cite{Numata2002}; good candidates for specific materials were Corning 7890-0F, and Heraeus Suprasil-312. The latter has a slightly lower bulk loss, but both make excellent candidates. Corning 7890-0F is used in the design example cases in Section \ref{sec:minther} as it is a material the authors have used to make high \ac{Q} resonators.
\par 
The roughness of samples produced by FEMTOprint was measured. The surface roughness was studied using a laser scanning microscope. Two regions of interest were measured: one where the femtosecond pulsed laser's angle of incidence was parallel to the remaining surface and one where it was normal to the remaining surface. The profiles and the average value taken were measured over nine lines in a hashed pattern. Ultimately, the parallel incidence side had a $R_{\rm{z}}$ (the mean Peak-Valley distance of 5 chunks of the sample region) of \SI[separate-uncertainty = true]{7.0(10)}{\micro\meter}, but a $R_{\rm{z}}$ of \SI[separate-uncertainty = true]{10.0(7)}{\micro\meter} was measured for the normal incidence beam, which is used as an estimate of $d_{\rm{s}}$ for the rest of this work. A cut-off wavelength of $\lambda_{\rm{c}}=$0.8\,mm was used for both these estimates.
\section{Design of Low Noise Mechanical resonators}
\label{sec:DesignofLow}

\begin{figure}
    \centering
    \begin{tikzpicture}[>=latex,scale=.75,decoration=snake,decoration=snake,every node/.style={scale=1}]
        \node[]at(4,1.5)(upperpoint){};
        \node[]at(-4,-1.5)(lowerpoint){};
        
        \node[]at(4,0.5)(upperpointp){};
        \node[]at(-4,-2.5)(lowerpointp){};
        
        \node[]at(-5,.875)(lowerpointpp){};
        \node[]at(3,3.875)(upperpointpp){};
        
        \node[]at(0,1.37)(centreofrot){};
        \node[]at(0,.37)(centreofrot2){};
        
        \draw[very thick](upperpoint.center)--(lowerpoint.center)node [midway](midwayNode){};
        \draw[very thick](upperpointp.center)--(lowerpointp.center);
        \draw[very thick](upperpointpp.center)--(lowerpointpp.center);
        \draw[very thick](upperpointpp.center)--(upperpoint.center)--(upperpointp.center);
        
        \begin{scope}[rotate around={20:(centreofrot)}]
           \draw[very thick, rounded corners](-1,0.37)rectangle(1,2.37);
           \node[]at(-1,0.37)(flexCorner1){};
           \node[]at(-1,2.37)(flexCorner2){};
           \node[]at(1,2.37)(flexCorner3){};
           \node[]at(0,.37)(flexwidthmarker){}; 
        \end{scope}
        
         \begin{scope}
            \path [clip,rotate around={20:(centreofrot)}] (-1,0.37)rectangle(1,2.37);
            \draw[very thick,  rounded corners,rotate around={20:(centreofrot2)}](-1,-.63)rectangle(1,1.37);
        \end{scope}

        \draw[triangle 45-triangle 45](upperpoint.east)--(upperpointp.east) node [align=left,midway, right] {Flexure\\ Height,\\ $d_{\rm{fheight}}$};

        \draw[decoration={markings,mark=at position 1 with
    {\arrow[scale=.5,>=triangle 45]{>}}},postaction={decorate}](flexwidthmarker.center)--(.45,3*.45/8);
    \draw[decoration={markings,mark=at position 1 with
    {\arrow[scale=.5,>=triangle 45]{>}}},postaction={decorate}](.45,3*.45/8)--(flexwidthmarker.center);
    \node[ align= justify,right,fill=white,rounded corners=2pt,inner sep=1pt, align= justify]at(.45,3*.45/8-.4)(fwidthlab){Flexure Width,
     $d_{\rm{fwidth}}$};
    
    \draw[triangle 45-triangle 45](flexCorner1.west)--(flexCorner2.west);
    
    \node [below,fill=white,rounded corners=2pt,inner sep=1pt,align=right]at(-3.1,1.1){Flexure Separation,\\
    $d_{\rm{fsep}}$};
    
    \draw[triangle 45-triangle 45](flexCorner3.north)--(flexCorner2.north)node[midway, above,align=left] {Flexure Length, \\$d_{\rm{flength}}$};
    
    \draw[-triangle 45](-3,3)--(flexCorner2.center);
    \node[align=right,left]at(-3,3){Inner Radius, \\ $R_{\rm{Inner}}$};
    
    \node[align=right,right]at(4,2.5)(testmass){Test Mass};
    \draw[-triangle 45](testmass.west)--(1.5,2.5);
  
    \path[](lowerpoint)--(lowerpointpp)node[midway](clamppoint){};
    \fill[](clamppoint)circle(.1);
    \node[below]at(clamppoint)(tom){Clamp Point};
    
    \end{tikzpicture}

    \caption{The geometric definition of the terms discussed in flexure design. Although the overall geometry of the device may change, the definitions of flexure length, separation, width and height stay the same. In this case one flexure junction is shown, in others, multiple flexure junctions may be used. The test mass is the suspended mass that is not rigidly attached to the ground, it is always on the opposite side of the flexure junction to the clamping point. The clamp point is the point (or interface) at which the monolithic piece is rigidly attached to the ground.  }
    \label{fig:paramdef}
\end{figure}
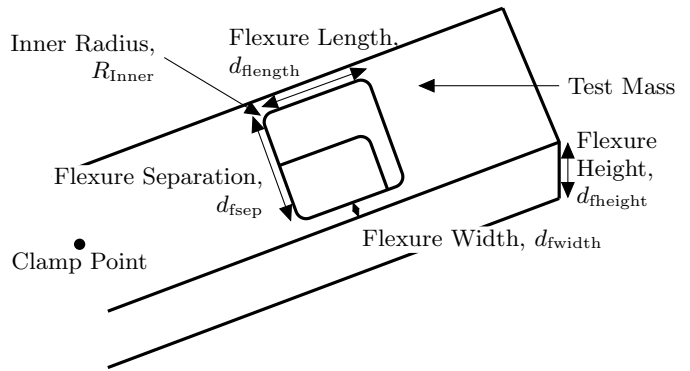
The flexures' behaviour is defined by its geometry, much more than the overall geometry of the resonator. Therefore, we wish to use simplified geometry to study good flexural design techniques. Well designed flexures can then be fit into an overall geometry based on the use case.
We consider a simple parallelogram linear geometry as shown in Figure \ref{fig:paramdef}. 
The parameters shown in Figure \ref{fig:paramdef} are studied throughout this section to show their effects on the resonator. The optimum flexure parameters are discussed through Section \ref{sec:minther}.
\par
\begin{figure}
    \centering
    \subfigure[Fundamental Mode]{\includegraphics[width=.2\textwidth]{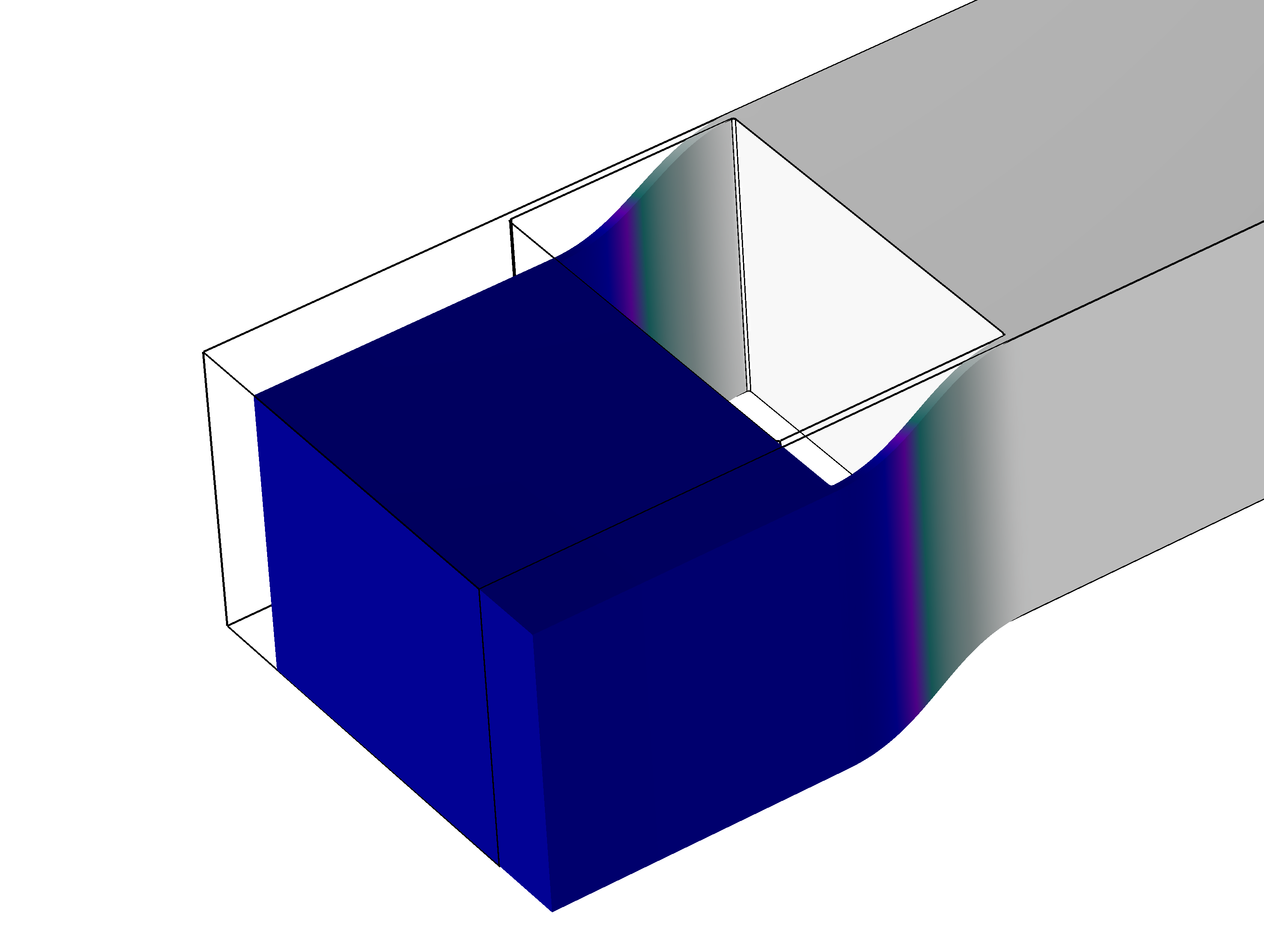}\hfill}
    \subfigure[Springboard Mode]{\includegraphics[width=.2\textwidth]{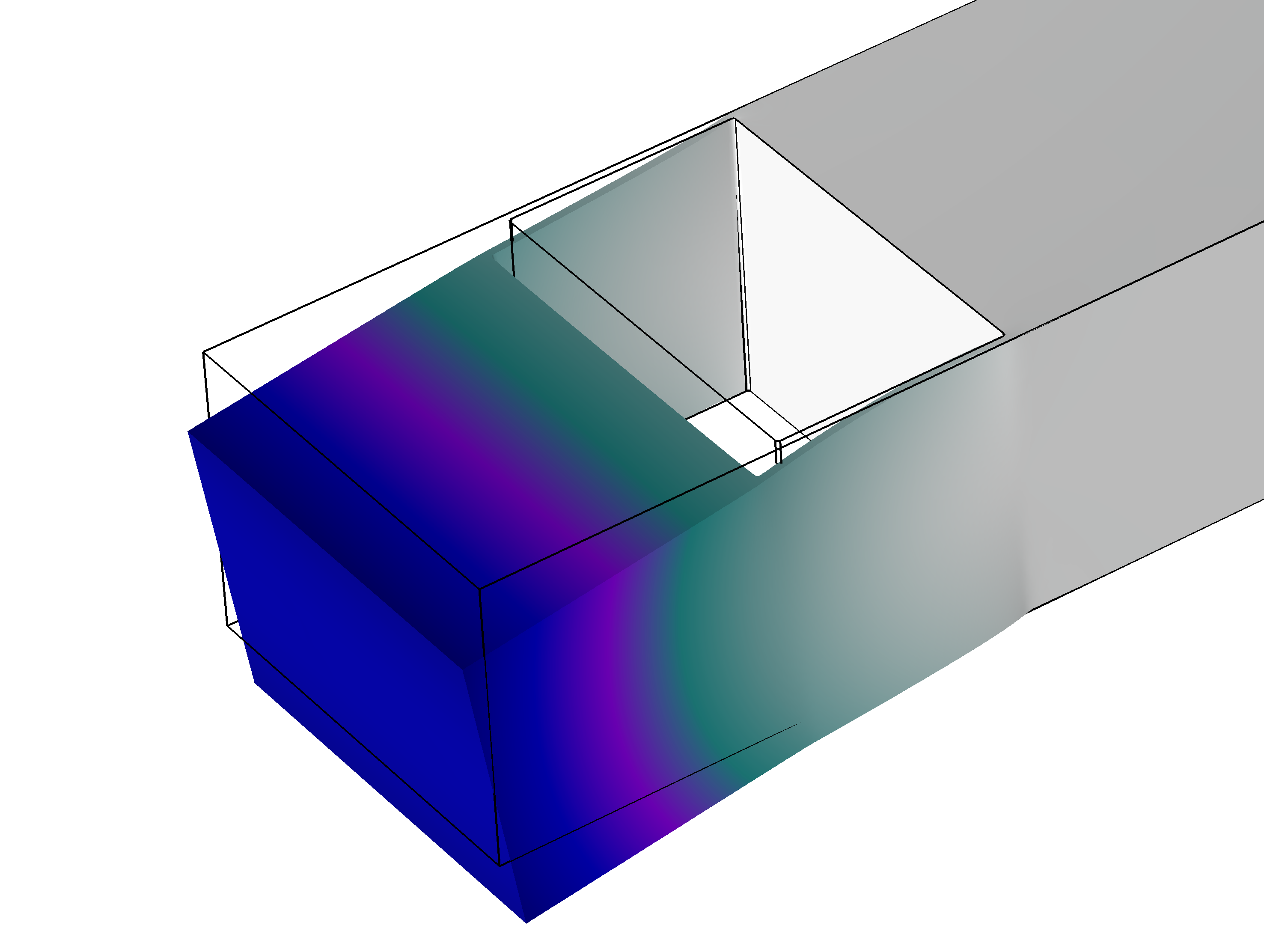}\hfill}
    \subfigure[Torsional Mode ]{\includegraphics[width=.2\textwidth]{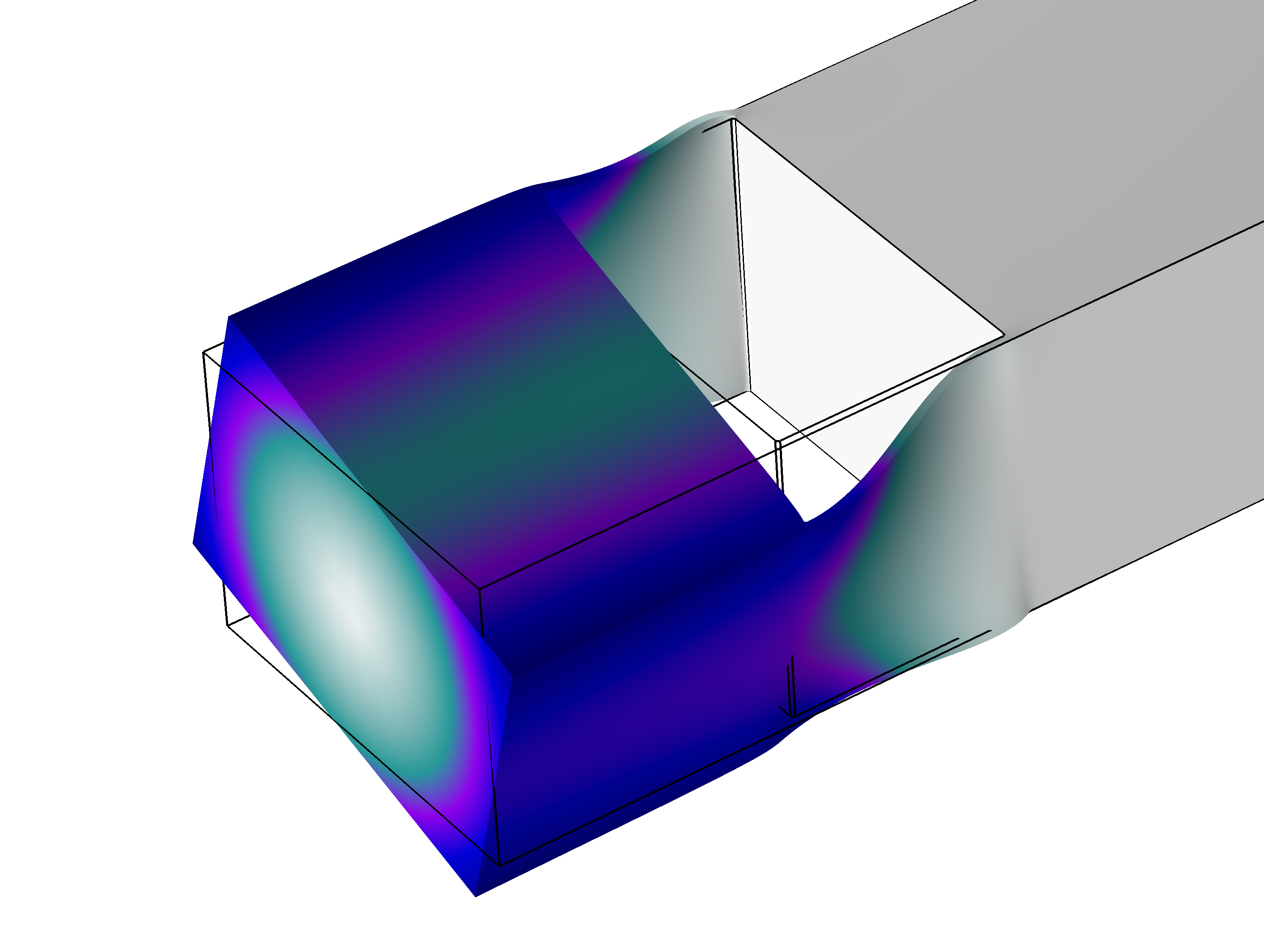}\hfill}

\caption{The first three eigenmodes for a linear parallelogram resonator. The fundamental mode is linear regarding excitations from the side, which can be used to track inertial motion. The springboard motion acts up-down, perpendicular to the direction of fundamental oscillation, but acts as an additional energy loss from the system. The torsional mode will couple directly to measurements of test mass displacement, and so must be suppressed using design. The heat map shows the local displacement at the extrema of the oscillation with arbitrary scaling. }
\label{fig:springboard}
\end{figure}

 Such geometries have been produced as sensors in several applications \cite{Guzman2014,Hines2020,Hines2023,Carter2022}. The parallelogram structure forces the fundamental mode into a linear in-plane motion, as shown in Figure \ref{fig:springboard}, which makes reconstructing inertial motion from measured test mass motion possible with simple transfer functions.
\par
The key elements we will study are how to tune the frequency of the fundamental mode, make the resonators survive in their intended operational environment, and combine these to optimise a design's thermal noise.
\subsection{Tuning resonator Resonances}
The most critical parameter in the design of an inertial sensor's resonator is its fundamental resonance frequency, $f_0$. $f_0$ defines how the system responds to external forces, determines what readout methods are appropriate both with respect to dynamic range and noise performance, and contributes to the thermal noise limitations. 
\par
Along with achieving a target $f_0$, we wish to tune other modes so that there is a significant frequency gap between them and $f_0$ and the target sensitive bandwidth. Various problematic effects can occur when these modes are too close, such as phonon-phonon loss \cite{Nawrodt2013}, beam pointing issues in optical readout schemes, and non-linear effective motion at the point used for displacement sensing.
Figure \ref{fig:springboard} shows the shape of the first three oscillation modes for a linear resonator. 
\par
In order to study tuning of mode frequencies COMSOL Multiphysics \ac{FEA} was used. The PARDISO solver was used to calculate solutions. For most simple geometries, the solution converged to a relative tolerance between iterations of $10^{-16}$ after ten iterations. Some more complex geometries required closer to 100 iterations to achieve this level of convergence. If the study was not converging, a finer mesh size was used. 
\par
\begin{figure*}
    \centering
    \includegraphics[width=\textwidth]{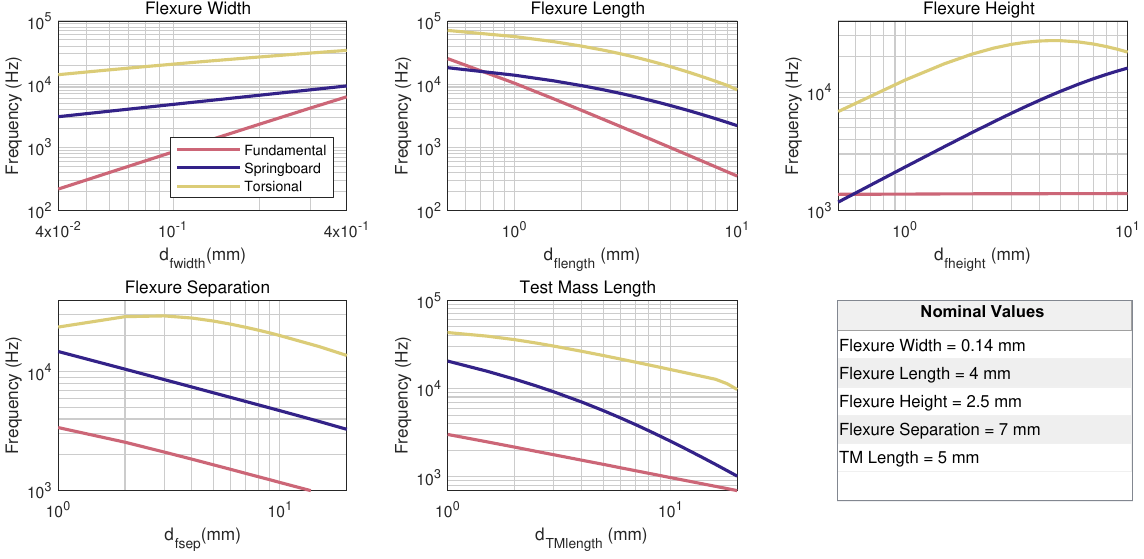}
    \caption{Graphs showing how the eigenfrequencies of a linear parallelogram resonator's first three modes of oscillation vary with changing parameters from Figure \ref{fig:paramdef}. The table of nominal values shows the parameter's value when it is not under investigation. The eigenfrequencies were predicted using COMSOL Multiphysics \ac{FEA}. For geometric reasons when the height or separation are increased the mass is also linearly increased.}
    \label{fig:params2modes}
\end{figure*}

A linear resonator was studied with the parameters shown in the last panel of Figure \ref{fig:params2modes}. For each graph, one parameter is varied while keeping the other four parameters constant. When not studied, they took the values shown in the last panel. The effect on each of the three fundamental modes is shown in the relevant plots.
\par
\par 
The fundamental mode evolves with respect to the geometric parameters as
\begin{equation}
        f_0\propto\sqrt{\frac{{d_{\rm{fwidth}}}^3{d_{\rm{fheight}}}}{{d_{\rm{flength}}}^3m}}.
        \label{eqn:f0}
\end{equation}
Parameters which add stiffness in the direction of the oscillation cause the frequency of that mode to increase by power 3/2, while adding general stiffness or mass increases it by power 1/2. 
\par
The other modes do not evolve as standard power laws. Many follow peak-shaped relations as different effects take over and dominate the response. 
In general, we see that increasing $d_{\rm{fheight}}$ best optimises the gap between modes, while lower frequency fundamental modes can be achieved both by reduction of ${d_{\rm{fwidth}}}$ or ${d_{\rm{flength}}}$.
\par
A problem one must be aware of is the internal modes in the flexures themselves. When too thin or long, the test mass and clamp side act as anchors and the system becomes akin to a system where a thin beam is clamped on both ends. This behaviour is especially problematic as the motion of the fundamental mode will directly couple to these modes and lead to a significant dissipation of energy from the system. 
\par
Furthermore, one must be careful of nonlinear effects from higher order modes coupling to the fundamental mode. The angle of the torsional mode, for example, would lead to tilt-to-length coupling \cite{Hartig2022} in the readout. The scale of this effect will depend on the specifics of the readout method and on how well the readout is centred on the test mass. Ultimately, the effect will lead to a limitation on dynamic range, as the measured motion will be dependent on the input forces to the sensors. In cases where precision readout is required in a noisy environment, this effect must be fully studied. In order to minimise this effect, a large frequency difference between the fundamental and higher order modes must be achieved, but this limit will depend on the use environment.
\par
Our FE analysis shows that the fundamental mode of oscillation can be separated from the springboard and torsional modes with a large flexure height and separation. Low frequency oscillation of the fundamental mode can be achieved by increasing flexure length and decreasing width.
\subsection{Resonator Survival}
\label{sec:LFO}
From section \ref{sec:nosie}, a need for low-frequency oscillations has been demonstrated. If the material is fixed in choice, this only leaves geometric considerations, which means reducing $d_{\rm{fwidth}}$ and $d_{\rm{fheight}}$, while increasing $d_{\rm{flength}}$ and increasing the mass of the suspended mass. Any of these changes make samples more prone to brittle fracture. 
\par
A simple failure model can be used if motion and stress are confined or largely dominated by one direction. In this case, the stress induced by a given load can be simulated using \ac{FEA}. Points in the geometry of high stress can be compared to the failure condition. The maximum shear stress criterion can be used to predict failure in resonators undergoing parallel motion, where most stress is confined to the direction of travel. We can then define a safety factor for all points in the sample 
\begin{equation}
    \frac{\textit{F}_{\rm{U}}}{\sigma}={\rm{Safety\ Factor}},
\end{equation}
where $\textit{F}_{\rm{U}}$ is the ultimate shear strength, and $\sigma$ is the shear stress at the given point in the resonator. A safety factor of less than 1 indicates the material will fail at that point; however, it is advisable to always leave some margin of error for extra unexpected loads. Furthermore, ultimate failure points in glass typically have large uncertainties as they strongly depend on local structure defects and contaminants.
\par
The highest load a resonator can expect to face depends on its environment. For many sensors used in a terrestrial environment in controlled laboratory space, the maximum stress induced is when gravity acts in the direction of oscillation and is free to fall. For this reason, along with the relatively high cost of making a sample, a safety factor of at least five whilst under 1\,g of load was used as a minimum baseline here.
\par
\ac{FEA} can be used to estimate the stresses across the material while under various loads. Again COMSOL Multiphysics was used to solve the stress distribution. The solid mechanics and heat transfer in solids packages were used to study the stress effects under prescribed test mass displacements. The displacement of a resonator under load from gravity is defined by 
\begin{equation}
    d_{\rm{sag}}=\frac{g}{\omega_{0}^2},
\end{equation}
so the displacement a resonator must survive under depends on the resonance frequency.
\par
\begin{figure}
    \centering
    \subfigure[]{\includegraphics{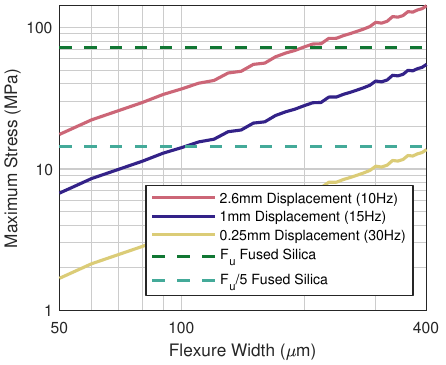}}
    \subfigure[]{\includegraphics{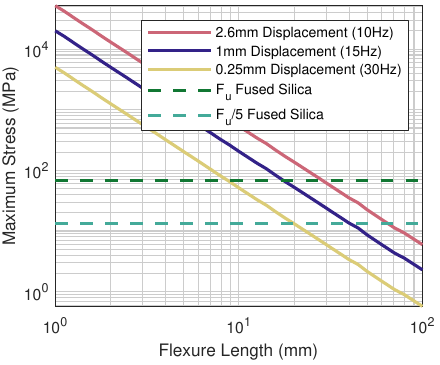}}
    \caption{The point of maximal stress in a fused silica beam with $d_{fheight} = 8$\,mm when undergoing flexural motion in the direction of the width, as a function of beam width (a) and length(b). When not being studied, the length is held at 40\,mm and the width at \SI{100}{\micro\meter}. The stress was simulated using COMSOL multiphysics. The solid lines correspond to different fixed displacements, which correspond to the sag related to a resonator of different natural frequencies under 1\,g load, shown in brackets. The dashed lines correspond to the ultimate shear strength of fused silica, $F_{\rm{u}}$ and this divided by five;  any resonator under this line has a safety factor of five when tilted against gravity.  }
    \label{fig:lengthstress}
\end{figure}

Several parameters can be tuned to increase the survivability of the resonators. One that may at first seem counter-intuitive is decreasing the flexure width. Naturally, one thinks of making the support thicker, and while this would lower the stress, it also increases the natural frequency. If one increases the test mass to compensate for this, the displacement remains the same, so the flexure must bend the same while being stiffer, increasing material stress. Therefore, thinning the flexure and supporting less weight will increase survivability. Practical considerations limit how thin the flexures can be manufactured, so we must look to other terms to improve survivability. Figure \ref{fig:lengthstress} (a) shows the effects of flexure width on maximal beam stress.
\par
Alternatively, the flexure length can be increased. Doing so reduces the curvature per unit length and the stresses induced. Figure \ref{fig:lengthstress} (b) shows the result of varying the beam's length; we can see the point at which a safety factor of five is reached. The Figure reasonably estimates the flexure required for a specific resonance frequency. A clear indication here is that, in this regime, the length requirement increases inversely proportional to the flexure length. Hence, the maximum flexure stress evolves as
\begin{equation}
    \sigma_{\rm{Max}}\propto\frac{{d_{\rm{fwidth}}}}{{d_{\rm{flength}}}^2f_0^2}.
    \label{eqn:stressscale}
\end{equation}
This result agrees with the behaviour of the analytically solved Timoshenko-Ehrenfest Beam Theory \cite{elishakoff2020handbook}, which describes beams free on one end. 
\par
The other parameter that can be tuned for optimal stress is the inner radii of the corners of the resonator. The gains from inner radii are limited but are studied in appendix \ref{app:ROC}.
\par
We find that to improve the survival of the oscillator against brittle fracture, the flexure thickness must be decreased and the length increased.

\subsection{Strain Distribution in Flexural Motion}
Equation \ref{eqn:bulkloss} shows how we need to understand the strain energy distribution through the flexure to estimate the \ac{Q} properly.
\par
The theoretical strain energy distribution was solved for an ideal, infinite plate with fixed edges by Norris \cite{Norris2005}. At the point of maximum displacement during oscillation, all energy should be potential energy. The strain energy distribution is given by 
\begin{equation}
    \mathcal{E}_{\rm{PE}}=\frac{{d_{\rm{fheight}}}^4}{24}\frac{1}{1-\nu^2}\left(E \kappa_{y y}^{2}\right),
\end{equation}
$\kappa_{{yy}}$ is the element of the curvature tensor with respect to the direction parallel to flexure length. Except for edge effects, the strain depends only on edge curvature and is uniform across the width and height. Therefore, we only consider effects across flexure length. If the flexure is homogenous across its length, the distribution can be estimated as uniform. The flexures can be designed so they are not homogenous along their length. We explore this in Appendix \ref{app:ROC}.
\section{Minimising Thermal Noise Through Flexure Design} 
\label{sec:minther}
We now use the information discussed so far to optimise the design parameters of the flexures in an inertial sensor. We consider two cases.
\subsection{Free Design}
\label{sec:freeDesign}
\begin{figure}
    \centering
    \includegraphics{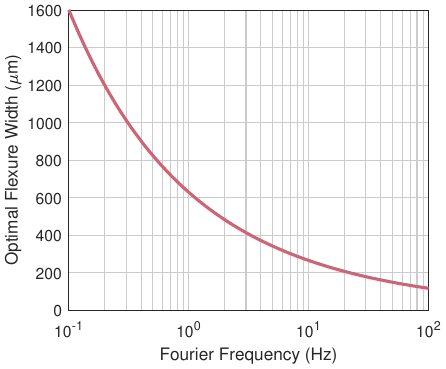}
    \caption{The flexure widths needed to achieve minimum thermal noise at a specific Fourier Frequency. Here, \ref{eqn:TNmincon} was minimised for flexure width for each Fourier Frequency, and the value was plotted. For this Figure, the approximation that strain density is uniformly distributed over flexure width is made (see Appendix \ref{app:ROC}), a material of Corning 7890-0F fused silica is used, with a surface loss of \SI{1e{-5}}{} \cite{Gretarsson1999}, with a surface depth of \SI{10}{\micro\meter} from measurements in Section \ref{sec:matloss}.}
    \label{fig:optimiseBeamWidths}
\end{figure}

\begin{figure}
    \centering
    \includegraphics{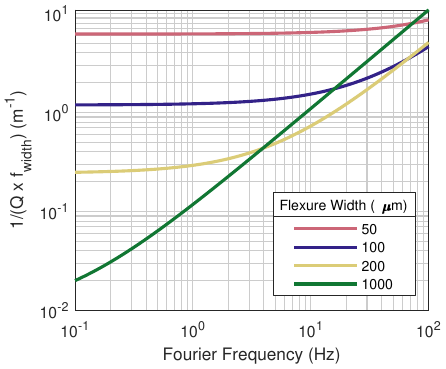}
    \caption{The results of the Term \ref{eqn:TNmincon} for flexures of 4 different widths across the Fourier frequency range of interest to the seismic isolation community. The strange units are effectively a proxy for thermal noise. The same assumptions, values, and material used in Figure \ref{fig:optimiseBeamWidths} are used here.}
    \label{fig:QFvolotuion}
\end{figure}

In the first case, we give an entirely free design space and aim to find the optimal design with respect to thermal noise. We will assume the sensor is operated in an environment such as a vacuum so that viscous damping is irrelevant and that the sensor aims to survive a defined load. We must consider \ref{eqn:strucNoise} to optimise structural damping. Each term in this equation is defined by the flexure parameters defined in Figure \ref{fig:paramdef}. 
The effects of each parameter were calculated and described in Section \ref{sec:DesignofLow}, and so the resonance frequency is related to the geometric factors by \ref{eqn:f0}.
Meanwhile, \ref{eqn:stressscale} shows the maximum stress in the flexure scales with length and thickness.
Combining these, the maximum possible mass for a given stress will evolve as 
\begin{equation}
    m\propto\frac{{d_{\rm{fwidth}}}^2\sigma_{\rm{Max}}{d_{\rm{fheight}}}}{{d_{\rm{flength}}}}.
    \label{eqn:meq}
\end{equation}
Substituting this into \ref{eqn:strucNoise}, we get that the acceleration thermal noise scales as
\begin{equation}
    	\tilde{A}_{\rm{TN}}\propto\sqrt{\frac{T}{{d_{\rm{flength}}}{Q(d_{\rm{fwidth}}){d_{\rm{fwidth}}}}{\sigma_{\rm{Max}}}^2}}.
    	\label{eqn:sTNopt}
\end{equation}
The ideal flexure width depends upon the target Fourier frequency in relation to the \ac{TED} peak and its magnitude with respect to surface losses. The evolution of the \ac{Q} with respect to $d_{\rm{fwidth}}$ is  
\begin{equation}
     Q=const \,\,\,\,\,\,\,\,\,\,\,\,\,\,\, (\rm{Bulk\,\,Loss\,\,Limited})
\end{equation}
\begin{equation}
     Q\propto  d_{\rm{fwidth}} \,\,\,\,\,\,\,\,\,\,\,\,\,\,\,(\rm{Surface\,\,Loss\,\,Limited})\end{equation}
     \begin{equation}
         Q\propto\frac{1}{{d_{\rm{fwidth}}}^2}  \,\,\,\,\,\,\,\,\,\,\,\,\,\,\,(\rm{TED\,\,Limited\,\,(Below\,\,Peak)})
     \end{equation}
     \begin{equation}
        Q\propto{{d_{\rm{fwidth}}}^2}  \,\,\,\,\,\,\,\,\,\,\,\,\,\,\,(\rm{TED\,\,Limited\,\,(Above \,\,Peak)}).  
     \end{equation}
These responses create two scenarios; the flexures should be tuned to be as thick as possible when above the \ac{TED} peak. Below the \ac{TED} peak, they should be tuned such that the term
\begin{equation}
        \frac{1}{{d_{\rm{fwidth}}}}\left({d_s}Q^{-1}_{\rm{Surf}}\frac{\int_{S} \epsilon(\vec{r}) \mathrm{d} S}{\int_{V} \epsilon(\vec{r})dV}+Q^{-1}_{\rm{Bulk}}+\frac { \alpha ^ { 2 } E T} {  C _ { \mathrm { p } } \pi^2\chi}\omega {d_{\rm{fwidth}}^2}\right),
        \label{eqn:TNmincon}
\end{equation}
is minimised across the desired frequency range, where we have taken the equations for $Q_{\rm{Mat}}$ and $Q_{\rm{TED}}$ from Section \ref{sec:nosie}. Typically, for precision inertial sensors for seismic measurement a bandwidth of 0.1-100\,Hz is relevant. The flexures must be centimetres thick for low-frequency seismic isolation to push the \ac{TED} peak below this band. Such thick flexures are not a feasible design strategy with current manufacturing techniques and go into ranges where assumptions we have made break down. Hence, we should focus on tuning the peak frequency to above detection bands. We consider how best to do this for a single flexure in Figure \ref{fig:optimiseBeamWidths}. We can use this graph to choose a flexure width for a sensor targeting a specific bandwidth. However, if we desire sensitivity over a larger bandwidth, we have to be more careful and look across the relevant frequency region. In this case, it is better to look at the effect of flexure width on \ref{eqn:TNmincon}, across the whole bandwidth. Figure \ref{fig:QFvolotuion} shows this for four different flexure widths. Although a thicker flexure improves the thermal noise floor, it does come at the cost of high-frequency performance. There are a few ways of deciding on flexure width from this Figure. A maximum acceptable noise floor across all frequencies could be targeted (particularly if readout noise is already known), and a flexure width that does not violate this at any frequency chosen. Alternatively, some (possibly weighted) average taken over the desired range could be used as a minimisation criterion. Often, this decision is also defined by what can be made. Ultimately, optimum flexure width is a very project-specific definition. Expression \ref{eqn:TNmincon} can be used to find a ${d_{\rm{fwidth}}}$ that meets these conditions for any targeted performance
\par
With two caveats, the flexure length should be tuned as long as possible. Lengthening the flexures is the most difficult parameter change to achieve with current manufacturing methods in terms of cost and complexity. Even if this can be achieved, there are still fundamental limits to how far flexure length can be taken. As the flexures get longer, the test mass gets smaller. Eventually, a point will be reached where internal flexure eigenmodes are significantly induced, disturbing the test mass position. 
\par
The flexure height and separation must be tuned to give the correct $f_0$ and $m$ from \ref{eqn:f0} and \ref{eqn:meq}. Ultimately, this will lead to lower heights and masses with softer springs and lower $f_0$ winning out in terms of thermal noise. Factors such as radius of curvatures were discussed in section \ref{app:ROC}, but already have fixed optimums without the need for trade-offs. A study following the steps discussed can find these points for any flexure, and the only reasons not to include them are for manufacturing simplicity. 
\subsection{Resonance Frequency Fixed}
\label{sec:RFF}
Often, limits on readout and application define $f_0$. With $f_0$ defined, the question is how to optimise around this restriction for thermal noise. Again, we can consider the same factors as Section \ref{sec:freeDesign}, but now with the additional criteria that $f_0$ is constant. The limitation places a restriction 
\begin{equation}
    {d_{\rm{flength}}}\propto \frac{d_{\rm{fwidth}} {d_{\rm{height}}}^{1/3}}{m^{1/3}}
    \label{eqn:lengthcon}
\end{equation}
in order to maintain the correct resonance frequency. The restriction also applies to the maximum suspendable test mass
\begin{equation}
    m\propto {\sigma_{\rm{Max}}}^{3/2} {d_{\rm{fwidth}}}^{3/2} {d_{\rm{height}}}.
\end{equation}
With the resonance frequency fixed, thermal noise is controlled by the $mQ$ product. Substituting limitations into Equation \ref{eqn:strucNoise}, we get the scaling
\begin{equation}
    	\tilde{A}_{\rm{TN}}\propto\sqrt{\frac{T}{{d_{\rm{fheight}}}{Q(d_{\rm{fwidth}}){d_{\rm{fwidth}}}^{3/2}}{\sigma_{\rm{Max}}}^{3/2}}}.
    	\label{eqn:sTNopt2}
\end{equation}
The condition on width is slightly adapted from the free design case
\begin{equation}
        \frac{1}{{d_{\rm{fwidth}}}^{3/2}}\left({d_s}Q^{-1}_{\rm{Surf}}\frac{\int_{S} \epsilon(\vec{r}) \mathrm{d} S}{\int_{V} \epsilon(\vec{r})dV}+Q^{-1}_{\rm{Bulk}}+\frac { \alpha ^ { 2 } E T} {  C _ { \mathrm { p } } \pi^2\chi}\omega {d_{\rm{fwidth}}^2}\right)
        \label{eqn:TNmincon2}
\end{equation}
which must again be minimised with equivalent Graphs to Figure \ref{fig:optimiseBeamWidths} and \ref{fig:QFvolotuion}. This again leads us to a similar optimisation study as Section \ref{sec:freeDesign} for flexure width. 
\par
A key difference here is the scaling with flexure height. Increasing the flexure height allows a greater mass to be suspended, making the flexures stiffer. This scaling will allow for a heavier test mass without compromising maximal stress. The limits here are again similar to the length scaling in the previous case. The increased height will lead to a larger volume etched, increasing the manufacturing cost and complexity of the devices. Furthermore, figure \ref{fig:params2modes} shows that the torsional mode will eventually become problematic as the height is increased. The length of the flexures is defined by what is needed to survive under load. The supported mass can then be tuned to give the correct $f_0$. 
\subsection{Practical Design of Gram Scale Resonators}
We have now evaluated the tools we have at our disposal to geometrically design gram scale resonators for inertial sensors. With this a general method of sensor design can be reached. 
\par
The first step is to define the sensor requirements and limitations. If specific readouts and noise performance are needed, $f_0$ often can be fixed. If the sensor is intended to operate in a noisy , a greater mode spacing between the fundamental and higher order modes is needed and a study on the required spacing would eb needed. Furthermore, understand its use environments determines the maximum load that a sensor needs to be able to survive. 
\par
From here specific geometric parameters can be tuned to achieve the desired performance. The flexure width can be optimised with results from \ref{eqn:TNmincon} and \ref{eqn:TNmincon2}. The length can then be tuned to survive under the expected load, and the height and mass tuned to give the required $f_0$. The flexure separation and overall geometry can then be tuned to meet the required oscillation mode separation.
With these scalings and manufacturing techniques, a ``bang for your buck" approach is reached, whereby even when optimised, there is a scaling financial cost to reaching a specific noise performance. Usually, the best approach is to estimate the noise floor from the readout and design the resonator to achieve high enough performance that the thermal noise no longer dominates at the relevant frequencies. 
\par
When an ideal resonator is designed, tolerances on design parameters should be considered. The relevant tolerances will be subject to both manufacturing methods and overall geometry. Appendix \ref{sec:toleandMismatch} discusses the specific case of resonators produced through subtractive manufacturing methods such as those present by Bellouard et al.\ \cite{Bellouard2012}
\section{Conclusions}
We have thoroughly explored the design of mechanical resonators for use in gram-scale precision inertial sensors.
A simple geometry has been used to study the effects of various design parameters on the resonator's behaviour. From this behaviour, optimal flexure parameters have been shown, for example, in design cases. The simple geometry and method here are easily adapted for more complex geometries. The guides and results presented here act as a base for designing such resonators for a wide range of applications. 
\par
A clear target bandwidth and sensitivity should be the starting point when designing such resonators. We can define an optimum flexure width, leading to the required length. Flexure height and test mass can be tuned for specific resonance frequencies and thermal noise floors. Understanding how these parameters interact takes a complicated interlinked multivariable problem and reduces it to a series of linear optimal point studies. Following this approach will lead to better resonator designs.
\par
Ultimately, better resonator designs will lead to better inertial sensors. These, in turn, offer an effective solution to the many experiments seeking to isolate residual disturbances from seismic motion. With this, we will be better able to answer many fundamental physical questions still open today. As the technology of gram-scale inertial sensors progresses, it will move outside the confines of specialised physics laboratories and into general public use and consumer production. From here, the potential applications of the technology are wide open.
\section*{Acknowledgments}
The authors would like to acknowledge advice and support from Gerhard Heinzel, Benno Willke, and Harald L\"uck and insightful discussions with Felipe Guzman.
\par
JJC, PB, and SMK acknowledge funding in the framework of the Max-Planck-Fraunhofer Kooperationsprojekt ``Glass Technologies for the Einstein Telescope (GT4ET)".
\par
OG was funded by the Deutsche Forschungsgemeinschaft (DFG, German Research Foundation) under Germany's Excellence Strategy—EXC 2121 “Quantum Universe”—390833306.

\appendix

\section{Optimal Radius of Curvatures}
\label{app:ROC}
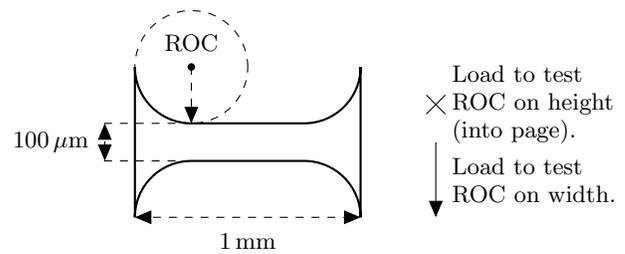
\begin{figure}
    \centering
    \begin{tikzpicture}[>=latex,scale=1,decoration=snake,decoration=snake,every node/.style={scale=1}]
        
        \draw[thick](0,-1)--(0,1)arc(180:270:.75)--(2.25,.25)arc(-90:0:.75)--(3,-1)arc(0:90:.75)--(.75,-.25)arc(90:180:.75);
        \draw[dashed](.75,1)circle(.75);
        \fill[](.75,1)circle(.05);
        \draw[dashed,-triangle 45](.75,1)--(.75,.25);
        \draw[](.75,1.1)node[above]{ROC};
        
        \draw[dashed](.75,.25)--(-.5,.25);
        \draw[dashed](.75,-.25)--(-.5,-.25);
        
        \draw[dashed,triangle 45 - triangle 45](-.4,.25)--(-.4,-.25);
        \draw[](-.5,0)node[left]{100\,$\rm{\mu}$m};
        
        \draw[dashed,triangle 45 - triangle 45](0,-1)--(3,-1);
        \draw[](1.5,-1.1)node[below]{1\,mm};
        
        \draw[-triangle 45](4,0)--(4,-1);
        \draw[align=left](4.1,-.5)node[right]{Load to test \\ROC on width.};
        
         \draw (4,.5) node[cross=4pt]{};
        \draw[align=left](4.1,.5)node[right]{Load to test \\ROC on height \\(into page).  };
    \end{tikzpicture}

    \caption{A diagram showing the flexure geometry used for the radius of curvature test. The design is \SI{100}{\micro\meter} deep into the page. A load can be applied normal to the geometry to study the effects of flexure height. To simulate a radius of curvature in the width the load can applied downwards as shown by the arrow.}
    \label{fig:rocdef}
\end{figure}

The effects of inner radius on stress distribution were studied by using \ac{FEA} simulation in COMSOL. The geometry shown in Figure \ref{fig:rocdef} was used for this test. A single flexure was tested, where a prescribed displacement of 25\,$\rm{\mu}$m was applied to one end and the other fixed in place, with no motion. The flexure was 1\,mm long and had a minimum width and height of \SI{100}{\micro\meter}. Then, the maximum stress along the flexures for different radii of curvatures was compared.
\par
\begin{figure}
    \centering
    \includegraphics[scale=1]{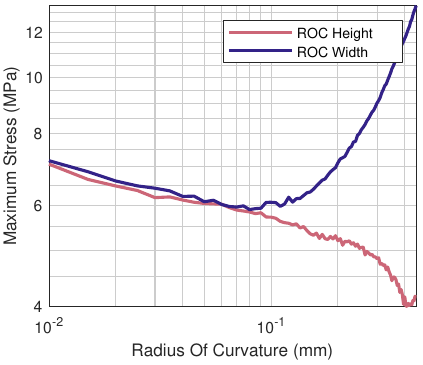}
    \caption{The Radius Of Curvature (ROC) vs.\ the maximal stress in a 1\,mm long sample undergoing flexural motion with a maximal end displacement of \SI{25}{\micro\metre}. The radius of curvature was applied in one of two directions, one in the direction of flexure width and one in the direction of flexure height. The minimum flexure width and height was \SI{100}{\micro\meter} when the curvature was applied in that direction. When the curvature was not applied in that direction, they were flat at \SI{100}{\micro\meter}. The simulation used COMSOL's stationary solver with its solid mechanics module.}
    \label{fig:rocStudy}
\end{figure}
Figure \ref{fig:rocStudy} shows the results of this simulation. A radius of curvature was applied in two directions. When the radius of curvature was applied to height, it was found that a larger radius distributes the stress better over the surface. Therefore, creating an hourglass reduces the maximum stress sustained in the flexure. Changing the width along the length of the flexure creates a minimum stress with a relatively small radius of curvature, which then increases again. Both these results have a simple physical interpretation. The width response results from a better stress distribution conflicting with thicker parts having to bend more. There is no effect of more material bending from extra height radius of curvature, so we only see the gains from better distribution. The small rise in stress as the flexure inner radius approaches half the flexure length is likely a result of stress from both ends of the flexure being pushed into the centre of the flexure. Therefore, the inner radius on the height should be designed just slightly short of the flexure length when manufacturing methods allow.  
\par
\begin{figure}
    \centering
    \includegraphics[]{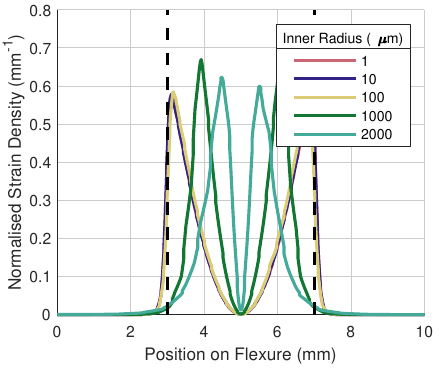}
        \caption{The normalised distribution of strain energy density for a cross-section across the flexures at positions along the flexure length. The energy is normalised, such that it is a ratio of total energy. This is shown for different inner radii of curvatures at the flexure ends. The black dashed lines show the ends of the flexure region and the return to the bulk material. The geometry studied is the same as defined in Figure \ref{fig:params2modes}}.
    \label{fig:straindis1}
\end{figure}

The other effect that must be considered with inner radii is the effect on surface loss. 
Since the material loss depends on the surface-to-volume ratio, adding more material to the flexures in high energy density regions makes sense, allowing better energy distribution along the flexure. The width can be altered by adding an inner radius to the flexure corner. 
A test was performed whereby different inner radii were tested to find an optimum point for strain distribution. The inner radius was added as indicated by Figure \ref{fig:paramdef}. Using a combination of static and eigenfrequency models, COMSOL was used to simulate these effects. The effect of different inner radii on strain distribution is shown in Figure \ref{fig:straindis1}. Adding the inner radius pushes the peak of the strain distribution inwards before decaying into the bulk. 
\par
\begin{figure}
    \centering
    \includegraphics[]{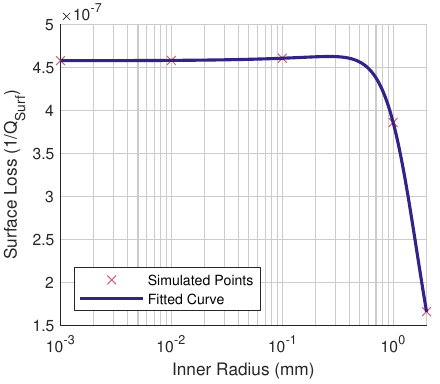}
    \caption{The surface loss of a resonator with geometry used in Figure \ref{fig:params2modes} when different inner radii are used. The simulation was done for the five inner radii in Figure \ref{fig:straindis1}, and fitted with a cubic interpolation routine. A larger inner radius leads to a lower surface loss.}
    \label{fig:surfLosscorn}
\end{figure}

The effect of this on surface loss is then shown in Figure \ref{fig:surfLosscorn}. Equation \ref{eqn:bulkloss} was used to calculate the surface loss, with the distribution of strain densities shown in Figure \ref{fig:straindis1} as the estimate for $\epsilon(\vec{r})$ from \ref{eqn:bulkloss}. The result of $d_{\rm{s}}$=\SI{10}{\micro\meter} is taken from \ref{sec:matloss} and a surface loss value of \SI{1e-5}{} from \cite{Gretarsson1999}. Figure \ref{fig:surfLosscorn} shows that a large inner radius leads to a lower surface loss. However, this only has a noticeable effect when the inner radii are above 10\,$\%$ of the flexure length. The corresponding study on extra stress in Figure \ref{fig:rocStudy} shows that stress becomes more localised and requires longer compensating flexures. The extra stress scales roughly as power 1/2, while the gain in surface loss scales roughly linearly. Considering the two cases discussed in Section \ref{sec:freeDesign}, we see that the gains of surface loss are cancelled out for the case $f_0$ is a free parameter above the optimum for any radius of curvature above its optimum stress point, Instead, if $f_0$ is fixed by external constraints, inner radii scales as power 1/4 with thermal noise. This is only true when surface losses are dominating. As the gains from increasing the inner radius are small, they should only be done when it does not add to manufacturing complexity. \par 
When $f_0$ can be chosen, adding an inner radius 10$\,\%$ of flexure length. Alternatively, when a specific $f_0$ is required, the optimum radius of curvature is a little under half the length, but we will only see gains while surface loss is limited. 
\section{Tolerances and Mismatches}
\label{sec:toleandMismatch}

The current discussion has so far focused on idealised cases. Once parts are manufactured, every defined parameter will have some deviation from the idealised case. The effect of these tolerances and how best to define them is discussed in this section. As the flexure region defines the resonator behaviour, and minor deformities significantly impact performance, tolerances on the flexures are the most important for resonator design. 
\par
The surface profiles of flexures produced using the methods discussed in this thesis are presented in Section \ref{sec:matloss}. The surface roughness is a measure of the very high spatial frequency effects. As such, its effects on the large-scale flexure geometry are limited and so only matter with regard to material loss calculations. The waviness profiles measured in this section are orders of magnitude lower than the quoted geometric deviations. The tolerances of concern are those that disturb the larger flexure geometry, especially in an uneven manner.  
\par
For this study, we will confine ourselves to low spatial frequency deviations from a specified flexure geometry. This would still be an infinite possibility space, but when consulting with companies and manufacturers, three typical deformations come to light that must be defined. Controlling the average width of the flexures across the structure is very difficult, with tolerances typically of the order of $\pm10\,\%$. It is also possible that the flexure width of one or both flexures varies along the length of the flexure. For example, if the wafer or etching machine is not perfectly flat, the width would change linearly over its length, creating a trapezoid geometry. The trapezoid geometry affects either or both flexures and could happen in the same direction or opposite to each other. Finally, the surfaces of the test mass and base facing each other may not be flush. As the distance between the two would vary, one flexure would have to be longer to compensate for this. We investigate the impact of each of these on the design throughout this section.
\par
After defining the parameters we wish to explore, we can study this again for the case of a linear resonator to understand the general effect of these mismatches. Similar studies would be needed on specific geometries that are to be manufactured. For example, we use the same linear resonator model from Figure \ref{fig:params2modes} and the same \ac{FEA} model. 
\par
\begin{figure}
    \centering
    \includegraphics[]{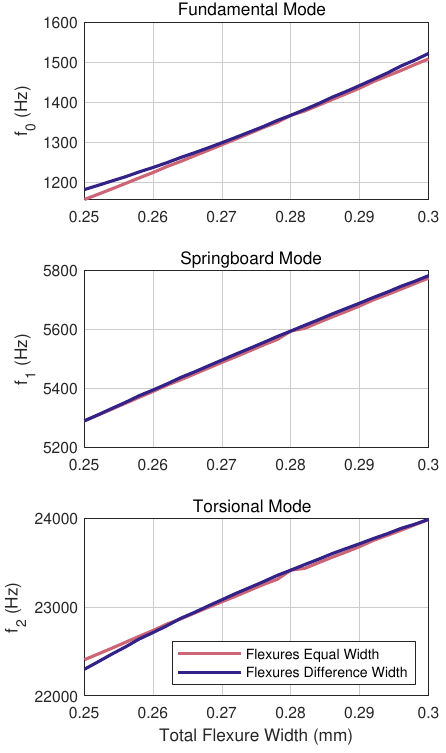}
    \caption{Dependency of the frequencies of the first three modes of a linear resonator, with parameters described in \ref{fig:params2modes}, with respect to flexure width mismatches. The graph shows two cases. The red curve describes when the total flexure width is evenly divided between the two flexures. On the other hand, the blue curve is when one flexure of the pair is fixed at 0.14\,mm thick while the other changes width. By comparing both, we see that the mode frequency is only slightly dependent on the ratio between the two flexures and is mostly dependent on the total width of the two flexures combined.}
    \label{fig:modesvsmismatch}
\end{figure}

The effect of varying average flexure width on the modes is shown in Figure \ref{fig:modesvsmismatch}. Here, two cases are studied, one where the average width is split evenly between the two flexures and one where one flexure is held at \SI{140}{\micro\meter} width and the other varied. We see limited effects on any mode from the distribution, but the mode frequency depends on the total flexure mass.
The only relevant mode frequency effect is if the device can survive operation with any flexure width within quoted manufacturing tolerances. This is best done by repeating \ac{FEA} simulations with extrema of the tolerances and checking safety factors in these cases.

\begin{figure*}
    \centering
    \includegraphics[]{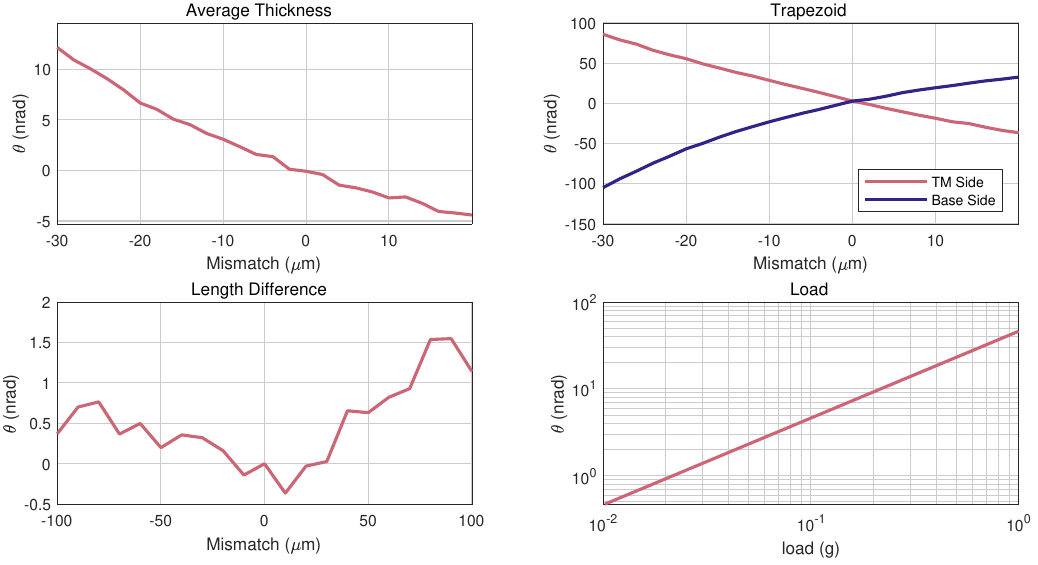}
    \caption{The angular dependence of a linear resonator with regards to a variety of tolerance mismatches between flexure pairs. The tests were conducted with a linear resonator as described in Figure \ref{fig:params2modes}. A stationary FEA analysis was performed in COMSOL. The tests were done with 1\,g of load on the resonator in the direction of the fundamental mode. The angle of the test mass relative to the load is shown for several cases. The first plot shows the case of two flexures, perfect cuboids with a width difference. The second graph shows when the flexure is a trapezoid, with the width changing over its length. The case where the base side has the correct width, and the test mass side is mismatched is shown in red, while the blue line shows the opposite orientation. The length difference shows the case of one flexure being longer or shorter than the other. The fourth plot shows how a fixed mismatch in average flexure width of \SI{20}{\micro\meter} scales with $\theta$ with differing loads.    }
    \label{fig:tolPlot}
\end{figure*}

Although the mode frequency effect was minimal in this case, the mode shape itself could be distorted. Imbalances between the flexures could lead to mode shape distortion. In practice, this distortion would mean the test mass no longer being perpendicular to the mode of oscillation. The effects of differing imbalance on test mass angle, $\theta$, relative to the load applied, were studied using the same linear resonator as the study in Figure \ref{fig:params2modes} and the results are shown in Figure \ref{fig:tolPlot}. These graphs show how $\theta$ evolves with mismatches between different parameters.
\par
Figure \ref{fig:tolPlot}(a) shows how under 1\,g load, $\theta$ varies with the changing width of one of two flexures. The simulation shows the expected result: when the flexures are the same width, there is no angular offset, but the angle increases as the flexures mismatch changes. $\theta$ scales linearly with respect to load, as shown in the fourth plot in this Figure. This is problematic as it is a noise source that scales with signal, making it challenging to model. The angle of the test mass will couple into readout measurement noise through \ac{TTL}. The effects of \ac{TTL} are subtle and numerous. They are thoroughly detailed by Hartig et al.\ \cite{Hartig2022}. The coupling to noise will vary enormously depending on the detector configuration and the readout scheme. With estimates of the angular noise introduced by the mismatch, the effect on the readout can be predicted using the models in the paper. 
\par

The effect of trapezoid-shaped flexures is shown in Figure \ref{fig:tolPlot}(b). Two cases are shown here, with opposite ends of the flexures being the ones that are mismatched. Both sides show similar results but with the rotation in opposite directions. When the base is varied, a slightly greater coupling to angle is seen, but both show a greater susceptibility to tilt than the purely average width change or length change cases. Tolerances should, therefore, be defined to minimise this criterion best. This is best done by defining a maximum and minimum value of the flexure width that cannot be exceeded at any point along the length. Typically, this criterion has an achievable $\pm$\SI{10}{\micro\meter} tolerance with the methods discussed by \cite{Bellouard2012}.
\par
The length of the two flexures was changed so that one was longer than the other. The result on $\theta$ is shown in Figure \ref{fig:tolPlot}(c). The effect of the length change was so small that uncertainties from the mesh elements seemed to dominate the test. Even with different lengths, each flexure can act independently and linearly without changing the mode shapes. Hence, this does not seem to be a critical criterion. 
\par
As the scale of this tilt coupling depends on load, as shown in Figure \ref{fig:tolPlot}(d), the relevance of this on the design will depend on environmental conditions.

\end{document}